\documentclass[iop]{emulateapj}
\usepackage{graphicx}
\usepackage{amssymb}
\usepackage{epstopdf}
\usepackage{float}
\usepackage{verbatim}
\usepackage{natbib}
\usepackage{amsmath}
\usepackage{mathptmx}
\usepackage[colorlinks=true,urlcolor=blue,citecolor=blue,linkcolor=blue]{hyperref}
\newcommand{\hi}{\ion{H}{1}}
\newcommand{\hii}{\ion{H}{2}}
\newcommand{\oi}{\ion{O}{1}}
\newcommand{\oii}{\ion{O}{2}}
\newcommand{\ci}{\ion{C}{1}}
\newcommand{\cii}{\ion{C}{2}}
\newcommand{\oil}{\oi~135.56$\,$nm line}

\newcommand{\iris}{\textit{IRIS}}
\newcommand{\bifrost}{{\textsl{Bifrost}}}
\def\kms{\mbox{km s$^{-1}$}}
\def\edt#1{{#1}}

\shorttitle{THE FORMATION OF THE \oi\ 135.56\,NM LINE}
\shortauthors{Lin \& Carlsson}

\begin{document}
\title{THE FORMATION OF \iris\ DIAGNOSTICS \\
    VII. THE FORMATION OF THE \oi\ 135.56\,nm LINE IN THE SOLAR ATMOSPHERE}


\author{Hsiao-Hsuan Lin$^1$}\email{h.h.lin@astro.uio.no}
\author{Mats Carlsson\altaffilmark{1}}\email{mats.carlsson@astro.uio.no}

\affil{$^1$ Institute of
 Theoretical Astrophysics, University of Oslo, P.O. Box 1029
 Blindern, N--0315 Oslo, Norway}

\begin{abstract}
The \oil\ is covered by NASA's {\it Interface Region Imaging Spectrograph} (\iris) small explorer mission 
which studies how the solar
atmosphere is energized.  We here study the formation and diagnostic potential of this line by means of non-LTE 
modelling employing both 1D semi-empirical and 3D radiation-Magneto Hydrodynamic (RMHD) models. 
We study the basic formation mechanisms and derive
a quintessential model atom that incorporates the essential atomic physics for the formation of the \oil. This
atomic model has 16 levels and describes recombination cascades through highly excited levels by effective
recombination rates. The 
ionization balance \oi/\oii\ is set by the hydrogen ionization balance through charge exchange reactions. The emission
in the \oil\ is dominated by a recombination cascade and the line is optically thin.
The Doppler shift of the maximum emission correlates strongly with the vertical velocity in its line forming region, 
which is typically located at 1.0 -- 1.5\,Mm height.
The total intensity of the line emission is correlated with the square of the electron density. 
Since the \oil\ is optically thin, the width of the emission line is a very good diagnostic of non-thermal velocities.
We conclude that the \oil\ is an excellent probe of the middle chromosphere, and compliments other powerful 
chromospheric diagnostics of \iris\ such as the \ion{Mg}{2} h \& k lines and the \cii\ lines around 133.5\,nm.

\end{abstract}


\section{Introduction}

 The \oi~135.56~nm intersystem line ($2s^2\,2p^3\,3s\,^5\!S_2 - 2s^2\,2p^4\,^3\!P_2 $) is covered by the NASA/SMEX mission {\it Interface Region Imaging Spectrograph} \citep[\iris,][]{2014SoPh..289.2733D}. 
Unlike other strong lines in the \iris\ spectral passbands, such as the h \& k resonance lines of singly ionised magnesium \citep{mg1, mg2, mg3} or the \cii\ 133.5 nm multiplet \citep{bhavna1, bhavna2}
that form in the upper chromosphere to lower transition region, the \oi\ 135.56 nm line generally forms at mid-chromospheric
heights, see below. 
Hence developing its potential diagnostics will provide a useful tool to  exploit the data from \iris\ and map different regions in the atmosphere.   
  
Already the SKYLAB mission discovered some  interesting behaviour of the \oil\ and its neighbouring permitted 
transition of \ci\ during a solar flare \citep{OICI-cheng80}. The \ci/\oi\ line ratio got greatly enhanced during the solar flare compared with 
its value in the quiet sun. The lack of a theoretical explanation makes it hard to interpret what the mechanism is behind such a dramatic change. 
\citet{OICI-cheng80} suggested that this line ratio can be sensitive to the chromospheric electron density, a critical parameter
for understanding the energy balance of chromospheric plasmas.
  
Non-Thermodynamic equilibrium (non-LTE) modelling of the neutral oxygen \edt{allowed resonance lines at 130.22, 130.49 and 130.60\,nm in the triplet system} and \edt{the} intercombination \edt{lines at 135.56 and 135.85\,nm} was carried out by \citet{oi1982}. 
The hydrogen Ly$\beta$ line is very close in wavelength to an oxygen resonance line \edt{in the triplet system} and radiative excitation of the oxygen line
by hydrogen Ly$\beta$ photons was found to be an important excitation mechanism for the \oi\ 130$\,$nm resonance \edt{lines}. 
The  excitation by Ly$\beta$ pumping is transferred to the lower excited states in the triplet system by radiative cascading. 
\citet{oi1982} also confirmed that the ionization of neutral oxygen is dominated by charge-exchange reactions with hydrogen and therefore 
is determined by the ionization degree of hydrogen. 
They only included one state ($2s^2\,2p^3\,3s\,^5\!S_2$) in the quintet system, which is the upper level of the intercombination \edt{lines}. 
Without the higher excited states in the quintet system, the authors concluded that the upper level of the \oil\ was mainly populated by 
collisional excitation. They also reported that the formation of the \oi\ \edt{intersystem lines} was less optically thick than the \oi\ resonance \edt{lines at 130\,nm}, 
due to the lack of a central reversal in the line profile.
    
The basic formation of the neutral oxygen lines was revisited by \cite{OI-carlsson92} with applications to a range of cool stars and solar 
chromospheric models, with a particular focus on the resonance \edt{lines at 130\,nm.}
The authors confirmed that the ionization degree of neutral oxygen is set by charge transfer with hydrogen, 
and the formation of the resonance  lines is dominated by the Ly$\beta$ pumping effect. 
The model atom developed by \cite{OI-carlsson92} was developed for the purpose of studying the resonance lines in the
triplet system. Since the ionization balance is determined by charge transfer reactions with hydrogen and the excitation is dominated
by hydrogen Ly$\beta$ pumping, highly excited states do not have to be included when the interest is focussed on the resonance
triplet. Their model atom has more excited states in the quintet system compared to \cite{oi1982} but the model atom is not complete
enough for the study of the intersystem lines that may be populated through a recombination cascade in the quintet system.

\citet{oi5409} studied a number of \oi\ lines for elemental abundance analysis, among them the \oi\ 777$\,$nm \edt{lines}, which belong
to the quintet system.  The upper levels of the 777$\,$nm lines are populated mostly through radiative cascading processes
within the quintet system, therefore such high-lying excited states were included in their study of the non-LTE effects on the oxygen abundance.
There was, however, no Ly$\beta$ pumping effects included in their modelling. Due to their inclusion of the highly excited levels 
in the quintet system, we use their model atom as a basis for constructing our simplified model atom.  

The layout of this paper is as follows: In Section~\ref{sec:method} we describe our radiative transfer computations and model atmospheres.
In Section~\ref{sec:quintessential} we construct a simplified model atom for the study of the \oil\ and describe this quintessential model atom.
In Sections~\ref{sec:steady_ion} and \ref{sec:steady_hpop} we discuss the basic formation mechanism and the effects from the hydrogen solution based on a 1D semi-empirical atmosphere. 
In Section~\ref{sec:steady_mov} we discuss different formation scenarios in a 3D atmosphere, in particular how velocity fields affect the line profile..
In Section~\ref{sec:diag} we present syntethic spectra and the potential diagnostics of the \oil.
In Section~\ref{sec:obs} we show the comparison of our simulation with observations, and we conclude in Section~\ref{sec:concl}.

 \section{Method}\label{sec:method}
 
 \subsection{Radiative transfer computations}
We use RH \citep{RH-uitenbroek01} to solve the non-LTE radiative transfer problem. 
RH is a multilevel accelerated lambda iteration code for radiative transfer calculations that includes 
partial frequency redistribution (PRD) treatment of the line profiles and also  blending between lines. 
We use the RH 1D version to study the basic formation mechanism in Section~\ref{sec:steady} with the FALC \citep{FALC} atmosphere. 
 For synthetic profiles from the 3D atmosphere, we solve the radiative transfer column-by-column as a 1D problem using the 1.5D version of 
RH \citep{Pereira:2015aa}. We take the non-LTE population densities of hydrogen from the model atmospheres and solve for
the non-LTE population densities of silicon (the element responsible for the dominant background opacity) in addition to oxygen.

 \subsection{Model atmospheres}
We use two model atmospheres for our study. In Section~\ref{sec:steady} we use the  semi-empirical solar chromosphere model 
FALC \citep{FALC} to study the basic formation mechanisms. The FALC model is a one-dimensional atmosphere in hydrostatic equilibrium that 
describes the atmosphere in an ill-determined averaged fashion. It is thus far from catching the dynamic state of the solar chromosphere but
serves the purpose of providing a test-ground with reasonable atmospheric parameters for studying atomic processes to develop an understanding
of the basic formation mechanisms. This understanding is essential for the development of a simplified atomic model that encompasses the 
essentials while being small enough to permit radiative transfer solutions in a 3D atmosphere.

To study the formation properties in a more realistic atmospheric model that catches some of the dynamic state of the solar 
chromosphere, we use a snapshot calculated with the 3D Radiation Magnetohydrodynamic (RMHD) code \bifrost\ \citep{Bifrost-code11}.
We use a  snapshot that extends 24x24x16.8 Mm$^3$ in physical space, with resolution of 504x504x496 grid-points. 
It covers the upper convection zone (from 2.5~Mm below the height where optical depth is unity at 500$\,$nm, the zero point of our height scale), 
the chromosphere, transition region and lower corona and includes magnetic field with two main patches 
of opposite polarity separated by 8$\,$Mm. 
The average unsigned field strength is 50G.  
The simulation cube is snapshot 385 of the {\it en024048\_hion} simulation that is available from
the European Hinode Science Data Centre
(http://www.sdc.uio.no/search/simulations). 
A detailed description of the simulation {\it en024048\_hion} is given in \citet{chromsim15}. 
This is the same simulation cube that has been used in the previous papers on the formation of \iris\ diagnostics
\citep{mg1, mg2,mg3,2015ApJ...806...14P}
and a number of other papers on line formation under solar chromospheric conditions
\citep{2012ApJ...749..136L, 2012ApJ...758L..43S, 2013ApJ...764L..11D}.

\section{Quintessential Model Atom}\label{sec:quintessential}

As a starting point for developing a simplified atomic model we use the 54-level model atom of \cite{oi5409}.
For the source of the atomic data, see that reference.
This 54-level model atom contains extra high lying excited states and updated collisional data compared with the 14-level model atom in \cite{OI-carlsson92}.   
We simplify the 54 level atom to a model atom with 16 levels to meet our purpose. We do this by replacing all levels
above the most highly excited level in \cite{OI-carlsson92} by effective recombination rates that catches the recombination cascade through these
levels without solving for their population density, see below.
Our resulting model atom is with 16 levels instead of 14 levels as the $ 2p^3 3p \ ^5P_1$, $ 2p^3 3p \ ^5P_2$ and $ 2p^3 3p \ ^5P_3$ are treated as separated levels, instead of a single merged term as in \cite{OI-carlsson92}. 
A major uncertainty in the model atom is the importance of collisions with neutral hydrogen. These collisions may couple the quintet system
with the triplet system and make the \oil\ emission sensitive to the pumping by hydrogen Ly$\beta$ radiation; a process that takes place
in the triplet system. 
Collisional data for collisions with neutral hydrogen are largely unknown.
A common approach in stellar abundance studies is to use a simplified recipe \citep{1968ZPhy..211..404D,1969ZPhy..225..483D} together
with a scaling factor determined from a fit to observations. This approach is \edt{not ideal} since a mismatch between observations and the 
non-LTE solution may come from the deficiency of other approximations (e.g., 1D plane-parallel, hydrostatic equilibrium atmospheres) and not
from lack of neutral hydrogen collisions. \cite{oi5409} included the Drawin recipe in their calculation but did not include the effect of Ly$\beta$
pumping. The effect on the \oi~777$\,$nm lines was rather small. We have made experiments using the 1D FALC model and find that 
neutral hydrogen collisions will make the Ly$\beta$ pumping contribute to the population of the upper level of the \oil\ and increase the 
intensity by up to 50\%. However, quantum mechanical calculations of neutral hydrogen collisional cross-sections for other elements
(none have been made for \oi\ yet) indicate that the Drawin recipe may overestimate the rates by several orders of magnitude 
\citep[see][for a discussion]{1993A&A...276..219C}. For this reason we have chosen not to include collisions with neutral hydrogen but
in light of the dependency we have found, it is important to recognize this source of uncertainty in the resulting intensities.
 
Highly excited levels are important for providing recombination channels from the continuum. 
After the recombination, we expect further downward cascading through allowed radiative transitions. 
In the chromosphere, we expect all these transitions to be optically thin such that we can neglect the reverse transitions (photoionization and
radiative excitation). 
This makes it possible to introduce effective recombination rates that take into account the recombination channels without treating the
levels in detail.

In general, we have a radiative recombination coefficient given by
 
\begin{equation}\label{eq:rci}
R_{ci} = 4\pi N_e\frac{g_{i}}{2g_{c}}(\frac{h^2}{2\pi m_{e} kT})^{\frac{3}{2}}e^{\frac{\chi_{ci}}{kT}}\int_{\nu_{0}}^{\infty} \frac{\sigma_{ic}(\nu)}{h\nu}(\frac{2h\nu^3}{c^2})e^{-\frac{h\nu}{kT}} d\nu
\end{equation}

where the index $c$ stands for the continuum, the index $i$ the level it recombines to, $Ne$ the electron density, $g_c$, $g_i$ the statistical weights, $m_e$ the electron mass, $T$ the temperature, $h$ the Planck constant, $k$ the Boltzmann constant, $\chi_{ci}$ the ionization energy, $\nu$ frequency, $\nu_{0}$ the threshold frequency, $\sigma_{ic}$ the photoionization crossection,  and $c$ the speed of light.

If we factor out the electron density, $Ne$, the rest of the expression in Eq. (\ref{eq:rci}) will be just a function of temperature after carrying out the integration over frequency. 
We can thus express the recombination to level $i$ as:

\begin{equation}\label{eq:rcit}
 R_{ci} = N_e \ R^*_{ci}(T)
\end{equation}

The advantage of such an approach is that if the net rate between the continuum and the state $i$ can be well approximated as this recombination
 process (meaning that the reverse photoionization is negligible), then we can replace the radiative recombination with a fixed rate that is
only a function of temperature. If the electrons in state $i$ further cascade down to other states with allowed transitions, e.g., to states $j$ and $k$, 
then the recombination to state $i$ can be split into recombination processes between the continuum and the states $j$ and $k$, 
with a split given by the transition probabilities for spontaneous deexcitation, $A_{ij}$ and $A_{ik}$:

\begin{equation}\label{eq:rci_split}
\begin{split}
 R^*_{ci}(T) &= \frac{A_{ij}}{A_{ij}+A_{ik}}R^*_{ci}(T) +  \frac{A_{ik}}{A_{ij} +A_{ik}}R^*_{ci}(T) \\
 &= R^*_{cj,i}(T) + R^*_{ck,i}(T)
\end{split}
\end{equation}

where $R^*_{cj,i}$ is the contribution to $R^*_{cj}$ from recombination through level $i$. This process can be repeated for all the levels where 
we want to take into account the recombination cascade without including the level in the model atom. For each level we keep in the model atom,
we thus end up with recombination contributions from all levels we exclude that couple to our level through allowed transitions. 
These contributions are just summed up to get the total effective recombination to that level.

\begin{figure*} 
   \centering
    \includegraphics[width= \textwidth]{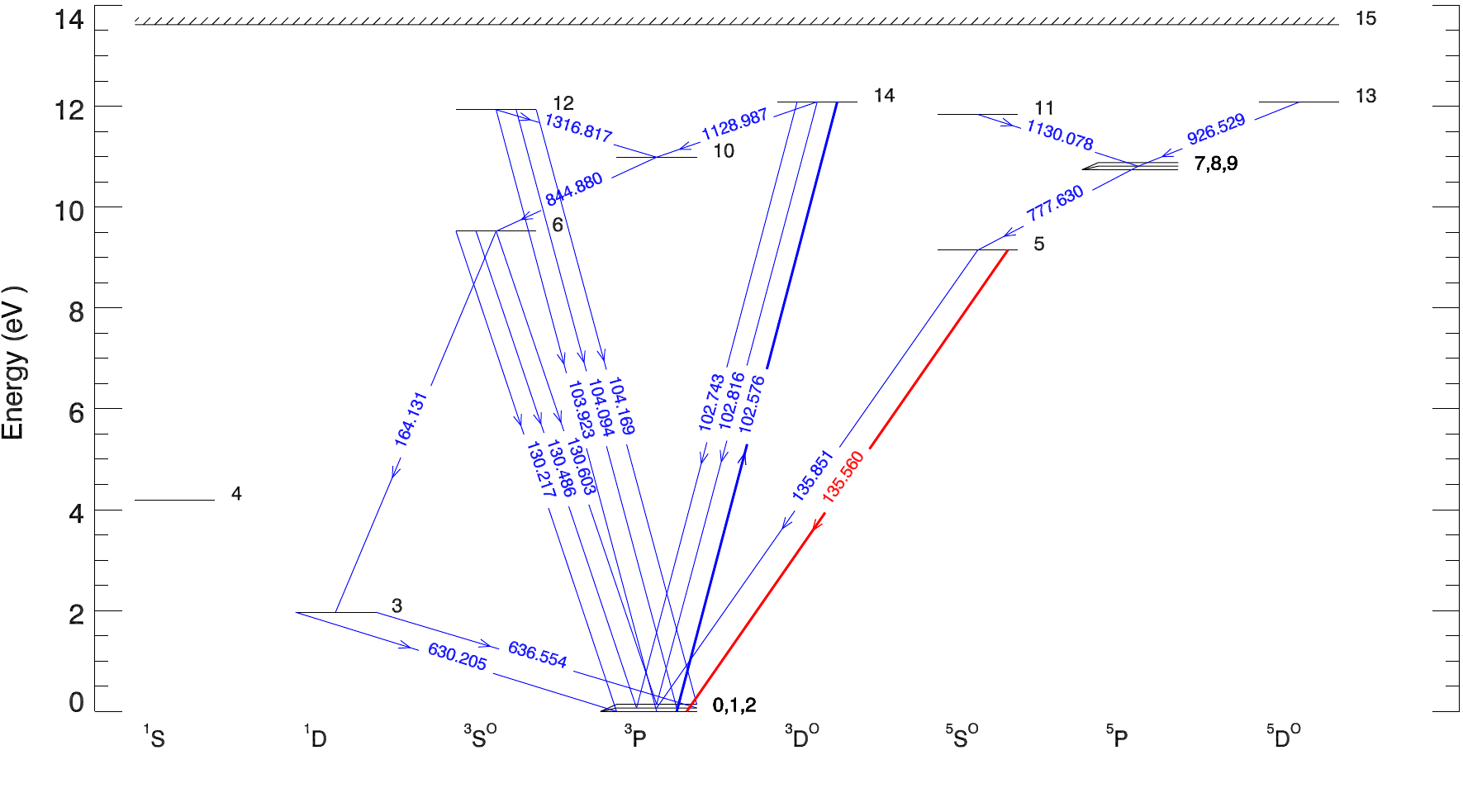}    	    	
   \caption{Term diagram of the 16 level \oi\ quintessential model atom used in this study. 
   The energy and designation of each numbered level is listed in table \ref{tb:oi_terms}. 
   The arrows show the direction of the net rates for each transition in the chromosphere. Vacuum wavelengths are shown in nm.
  The energies of the levels of the terms $2p^4\,^3\!P$ (levels 0,1,2) and $2p^3\,3p\,^5\!P$  (levels 7,8,9) are very close to each other 
  and the energy differences within these terms are exaggerated in this diagram.
  The transitions to or from levels 7,8 and 9 are represented by the transition to and from level 8 ($ 2p^3\,3p\,^5\!P_1$) only. 
  The \oil\ that is the topic of this paper is shown in thick red, the 102.576\,nm line that is pumped by the hydrogen
  Ly$\beta$ line is in thick blue.}
   \label{fig:o_term_diag}
\end{figure*}

\begin{figure} 
   \centering
    \includegraphics[width= \columnwidth]{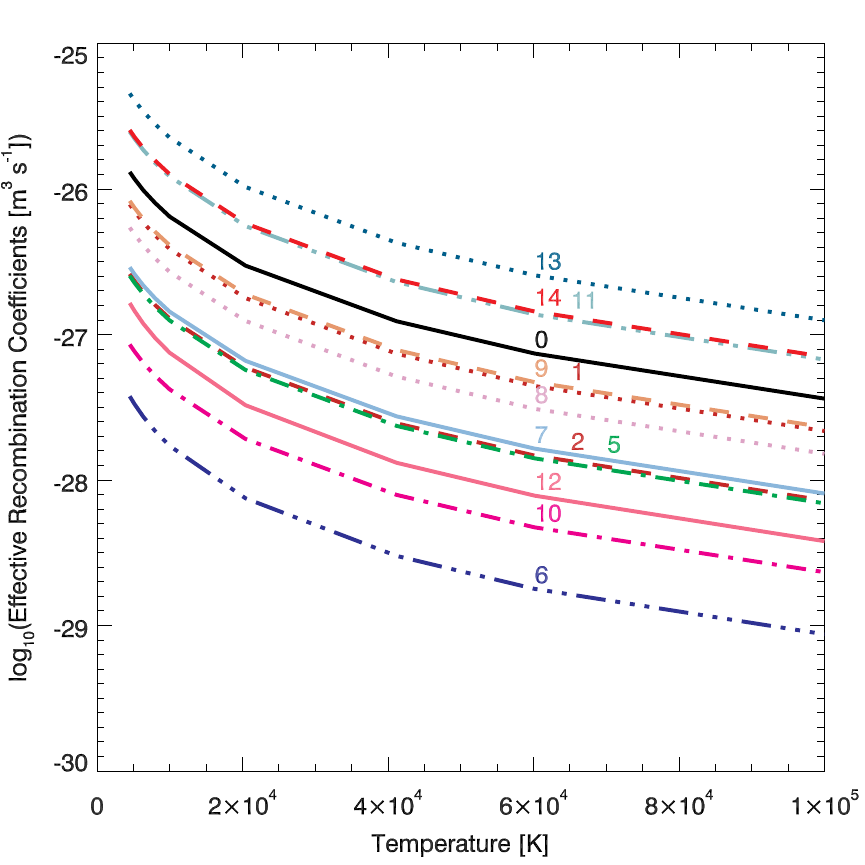}    	    	
   \caption{Effective recombination coefficients, see Table~\ref{tb:cr}. The labels of the excited states are on top of each line with the same color.
    If two lines are very close, the label of the line with slightly lower value is shifted toward the right.}
   \label{fig:ratecoef}
\end{figure}

Figure \ref{fig:o_term_diag} shows the term diagram of our quintessential model atom with 16 levels. The level energies and 
designations are given in Table~\ref{tb:oi_terms}. 
The effective recombination coefficients to all levels are tabulated in Table~\ref{tb:cr} and we show them in Figure~\ref{fig:ratecoef}.

\begin{table}
\centering
\caption{Atomic parameters for O.\label{tb:oi_terms}}
\begin{tabular}{rrc}
\hline\hline
Level             & Energy [$cm^{-1}$]& Designation \\
\hline
0                 &0.000 & \oi\ $2p^4\,^3\!P_2$ \\
1           &158.265  & \oi\ $2p^4\,^3\!P_1$\\
2          &226.977& \oi\ $2p^4\,^3\!P_0$\\
3          &15867.862& \oi\  $2p^4\,^1\!D_2$ \\
4          &33792.583& \oi\  $2p^4\,^1\!S_0$ \\
5          &73768.200& \oi\  $2p^3 \edt{3}s\,^5\!S_2^o$ \\
6          &76794.978& \oi\  $2p^3 3s\,^3\!S_1^o$ \\
7          &86625.757& \oi\  $2p^3 3p\,^5\!P_1$ \\
8          &86627.778& \oi\  $2p^3 3p \,^5\!P_2$ \\
9          &86631.454& \oi\  $2p^3 3p\,^5\!P_3$ \\
10          &88630.977& \oi\  $2p^3 3p\,^3\!P$ \\
11          &95476.728& \oi\  $2p^3 4s\,^5\!S_2^o$ \\
12          &96225.049& \oi\  $2p^3 4s\,^3\!S_1^o$ \\
13          &97420.748& \oi\  $2p^3 3d\,^5\!D^o$ \\
14          &97488.476& \oi\  $2p^3 3d\,^3\!D^o$ \\
15          &109837.020& \oii\   ground \\
\hline\hline
\end{tabular}
\end{table}

\begin{table*}
\caption{Effective Recombination Coefficients\label{tb:cr}} 
\begin{center}
\begin{tabular}{l*{9}{c}r}
\hline\hline
&\multicolumn{8}{c}{Temperature [K]}\\
      $R^*_{ci}$   & 4500&5160&6370&7970&9983&20420&41180 &60170&100000 \\
\hline
$R^*_{c14}$ [$10^{-28} \ m^3 s^{-1}$]&254.126&227.679&190.951&156.831&126.345&57.9456&23.8875&14.3800&7.03006 \\
$R^*_{c13}$ [$10^{-27}\ m^3 s^{-1}$]&45.3115&40.6429&34.0479&27.9286&22.5443&10.3130&4.26601&2.55122&1.25207 \\
$R^*_{c12}$ [$10^{-29}\ m^3 s^{-1}$]&164.436&145.251&118.711&94.9460&74.9470&32.6631&13.1234&7.80194&3.80117 \\
$R^*_{c11}$ [$10^{-28}\ m^3 s^{-1}$] &247.508&221.465&185.268&151.301&121.811&55.5946&22.9252&13.7275&6.72938\\
$R^*_{c10}$ [$10^{-29}\ m^3 s^{-1}$] &85.4289&76.4623&63.7694&52.0674&41.9316&19.1376&7.90372&4.72786&2.32700\\
$R^*_{c9}$  [$10^{-28}\ m^3 s^{-1}$] &82.8493&74.1332&62.0482& 50.8526&41.0086&18.8720&7.82051&4.69823&2.30519\\
$R^*_{c8}$  [$10^{-28}\ m^3 s^{-1}$] &54.3390&48.6406&40.7773&33.3889&26.8978&12.3791&5.13489&3.08086&1.51400\\
$R^*_{c 7}$  [$10^{-29}\ m^3 s^{-1}$] &289.806&259.521&217.337&177.804&143.389&65.9316&27.3822&16.4199&8.07420\\
$R^*_{c6}$  [$10^{-30}\ m^3 s^{-1}$] &376.007&332.290&272.020&217.439&171.643&74.8017&30.0632&17.8556&8.70174\\
$R^*_{c5}$  [$10^{-29}\ m^3 s^{-1}$] &253.012&226.401&189.665&154.888&124.757&57.0020&23.5347&14.0899&6.90617\\
$R^*_{c2}$  [$10^{-29}\ m^3 s^{-1}$] &260.786&233.351&195.260&159.762&128.709&59.2080&24.5266&14.7083&7.23052\\
$R^*_{c1}$  [$10^{-28}\ m^3 s^{-1}$] &78.3535&70.0987&58.6938&47.9973&38.7065&17.7829&7.36847&4.42349&2.17186\\
$R^*_{c0}$  [$10^{-28}\ m^3 s^{-1}$] &130.846&117.275&98.0652&80.2916&64.5800&29.7279&12.3206&7.39798&3.62685\\
\hline\hline
\end{tabular}
\end{center}
\end{table*}

\section{Basic formation mechanism}\label{sec:steady}

\subsection{Ionization balance and the population of the upper level}\label{sec:steady_ion}

The ionization degree of the neutral oxygen is mainly decided collisionally through charge transfer with H and $\rm H^{+}$ \citep{judge86}, 

\begin{equation}
\rm O + H^{+} \leftrightarrows O^{+} + H
\end{equation}

and the relation between the oxygen populations and the hydrogen populations follows Eq. (3) in \citet{OI-carlsson92}:
\begin{equation}\label{eq:oion}
\frac{n(\mbox{\oi})}{n(\mbox{\oii})}  = \frac{9}{8} \frac{n(\mbox{\hi})}{n(\mbox{\hii})}.
\end{equation}

\begin{figure}
   \includegraphics[width= \columnwidth]{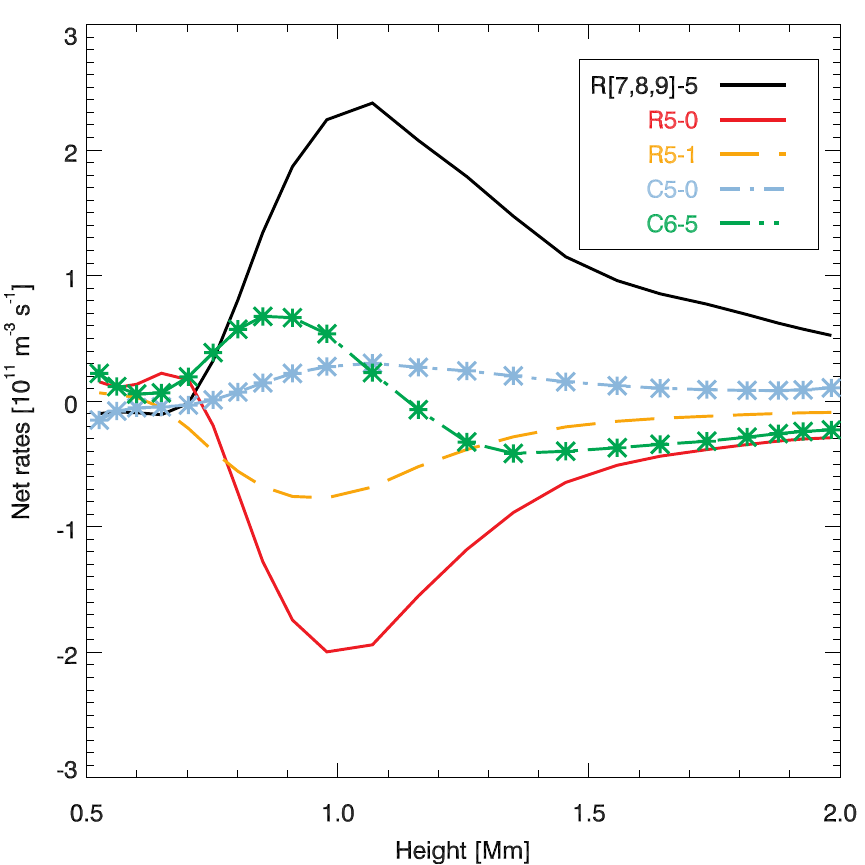}   
   \caption{Net rates to  $2p^3\,3s\,^5\!S_2$, 
   (level 5 in the \oi\ model atom (Figure~\ref{fig:o_term_diag})). 
   Positive values denote a net rate coming into the level, negative values a rate going out of the level.
   Letters R and C in the label stands for radiative net rate and collisional net rate, respectively. 
   The collisional net rates are also overlaid with star symbols.   
   The dominant channel into the $2p^3\,3s\,^5\!S_2$ level is from the 
three levels in the  $2p^3\,3p\,^5\!P$ term, (levels 7,8,9, solid black).
The secondary channel is through collisions from the $2p^3\,3s\,^3\!S_1^o$ 
term (level 6, green dot-dot-dashed with stars).
The $2p^3\,3s\,^5\!S_2$ level is depopulated by the emission lines at 135.56$\,$nm (solid red) and 135.85$\,$nm (dashed yellow).
There is also a collisional rate, C5-0, corresponding to the 135.56\,nm line (dot-dash blue with stars).
   }
   \label{fig:rate_o}
\end{figure}

The excited states in our atomic model belong to two main systems, the triplet and the quintet system. 
The upper level of the 135.56 nm line, $2p^3\,3s\,^5\!S_2$, belongs to the quintet system. 
The lower level of this line,  $2p^4\,^3\!P_2$, which is in the ground term, belongs to the triplet system. 
The 135.56 nm line is hence an intersystem line with low oscillation strength and therefore low opacity.
The population and depopulation of the upper level is shown in  Figure~\ref{fig:rate_o}.
The dominant channel for electrons to come into the $2p^3\,3s\,^5\!S_2$ level is from the 
three levels in the  $2p^3\,3p\,^5\!P$ term, (levels 7,8,9).
The secondary channel of incoming electrons is through collisions from the $2p^3\,3s\,^3\!S_1^o$ 
term (level 6), which is part of the triplet system. 
The $2p^3\,3s\,^5\!S_2$ level is depopulated by the emission lines at 135.56$\,$nm and 135.85$\,$nm,
which correspond to radiative rates R5-0, R5-1. 
There is also a collisional rate, C5-0, corresponding to the 135.56\,nm line, but its effect is rather small 
compared to the other channels.

The upper level of the \oil\ is thus populated mostly by radiative cascading from the continuum 
through the $2p^3\,3p\,^5\!P$ term and is depopulated through the 135.56\,nm and the 135.85\,nm line. 
As we will see in Section~\ref{sec:contri}, due to the low oscillator strength of our line, 
it is usually optically thin.

\subsection{Effects of hydrogen}\label{sec:steady_hpop}

The effects from hydrogen is twofold: one from the ionization degree of hydrogen, 
and the other from the radiation field of the Ly$\beta$ line. 

As a primary channel, the states in the quintet system will be populated by a cascading process from
the continuum and their population will therefore depend on the oxygen ionization degree. 
Since the oxygen ionization is dominated by charge transfer with neutral hydrogen and protons,  
higher hydrogen ionization degree will result in higher oxygen ionization degree 
(see Equation~\ref{eq:oion}). 

\begin{figure}
   \centering
    \includegraphics[width= \columnwidth]{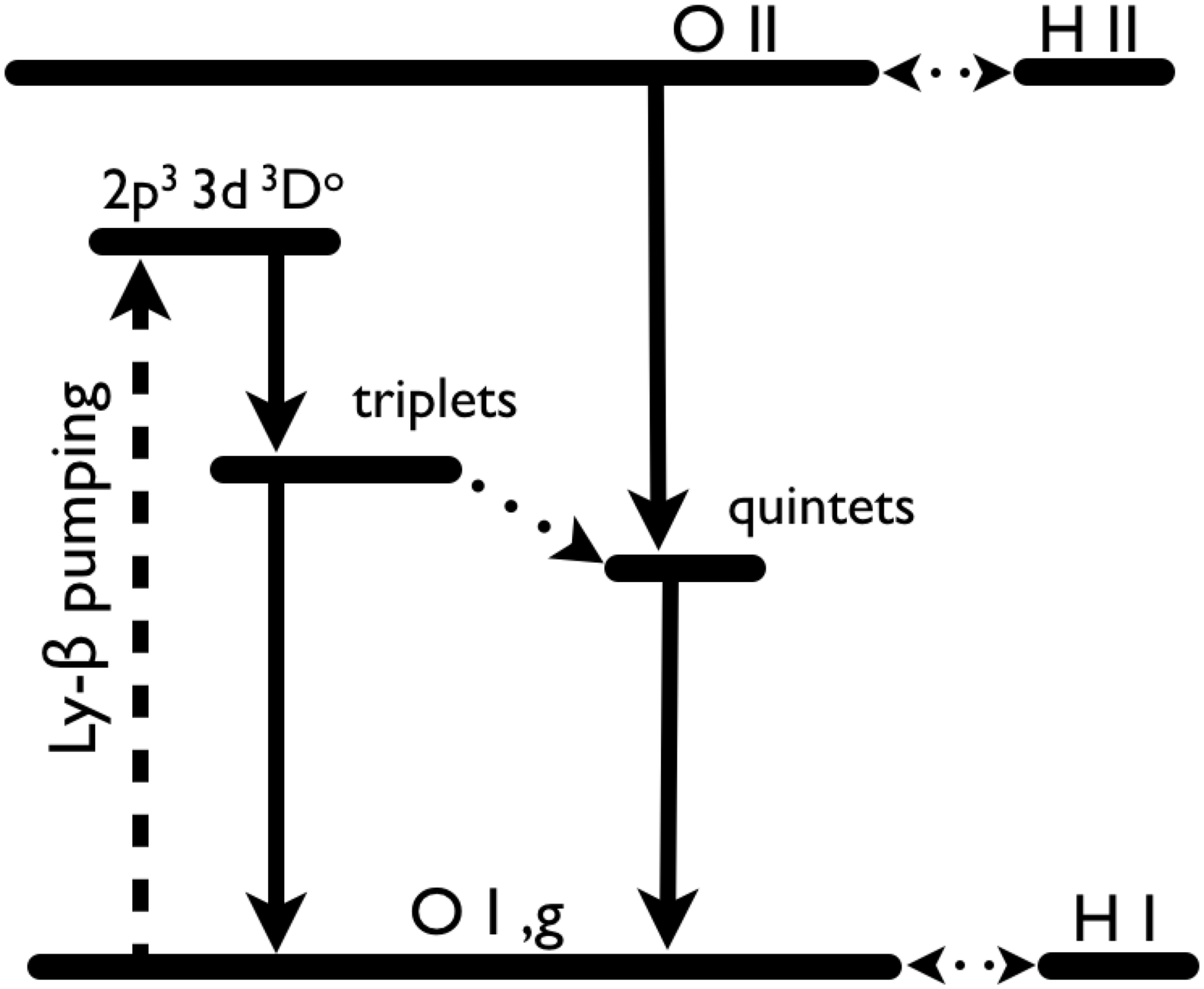}

   \caption{\oi\ term diagram that summarizes the basic formation of the \oi\ 
135.56 nm line (Section~\ref{sec:steady}) . 
The solid lines represent radiative cascading, the dashed line radiative excitation and
the dotted line stands for collisions. 
The ionization degree of \oi\ is set by the ionization degree of hydrogen through charge transfer. 
The main rate into the quintet system, where the state $ 2p^3\,2s\,^5S_2$ belongs, 
is through radiative cascading. 
Within the triplet system, the Ly$\beta$ pumping drives photo-excitation which is followed by 
radiative cascading within the triplet system. 
There may, however, also be a channel from the triplet system into the quintet system through 
collisions, thus making the \oil\ sensitive to Ly$\beta$ pumping.}
   \label{fig:o_term}
\end{figure}

As a secondary channel the quintet levels can also be populated through collisions
from the triplet levels (C6-5 in Fig.\ref{fig:rate_o}), and with such a channel the Ly$\beta$ 
radiation might have an effect on the 135.56\,nm emission as well.  
The reason is that the Ly$\beta$  pumping of the \oi\ 102.5 nm resonance line dominates the 
excitations in the triplet system \citep{OI-carlsson92}. 
We have made tests with and without the effect of
Ly$\beta$ pumping included and find negligible effect on the \oil\ if only collisions with electrons 
couple the triplet and quintet system. Collisions with neutral hydrogen may increase the coupling, but
the collisional cross-sections are unknown and likely not large enough to affect the \oil, see
Section~\ref{sec:quintessential} for a discussion. The main ionization/excitation and de-excitation 
channels are summarized in Figure~\ref{fig:o_term}.

\subsection{Line Formation in a 3D atmosphere}\label{sec:steady_mov}

We here illustrate the formation of the \oil\ by showing the formation in detail with the help
of the four-panel diagrams introduced by \citet{4panel}. 

The emergent intensity along the normal in a 1D plane parallel semi-infinite atmosphere can be written as
\begin{equation}\label{int}
I_{\nu} =\int_0^\infty S_{\nu} e^{-\tau_{\nu}} \chi_{\nu} dz
\end{equation}

where the source function ($S_{\nu}$), opacity ($\chi_{\nu}$) and 
optical depth ($\tau_{\nu}$) are functions of frequency ($\nu$) and geometrical height ({\em{z}}). 
The integrand describes the local creation of photons ($S_\nu \chi_\nu dz$) and the fraction of those
that escape ($e^{-\tau_\nu}$). 
We therefore define the contribution function to intensity on a geometrical height scale as
\begin{equation}\label{eq:cinu}
C_{I_{\nu}}(z)=S_{\nu} e^{-\tau_{\nu}} \chi_{\nu}.
\end{equation}

Following \citet{4panel} we can rewrite the contribution function as

\begin{equation}\label{eq:cisplit}
C_{I_{\nu}}(z)=S_{\nu} \tau_\nu e^{-\tau_{\nu}} {\chi_{\nu} \over \tau_\nu}.
\end{equation}

where the term $\tau_\nu e^{-\tau_{\nu}}$ has a maximum at $\tau_\nu=1$ and represents the Eddington-Barbier
part of the contribution function, $S_\nu$ gives the source function contribution and the final term,
${\chi_{\nu} \over \tau_\nu}$ picks out effects of velocity gradients in the atmosphere.
 
\begin{figure}  
   \centering
    \includegraphics[width=\columnwidth]{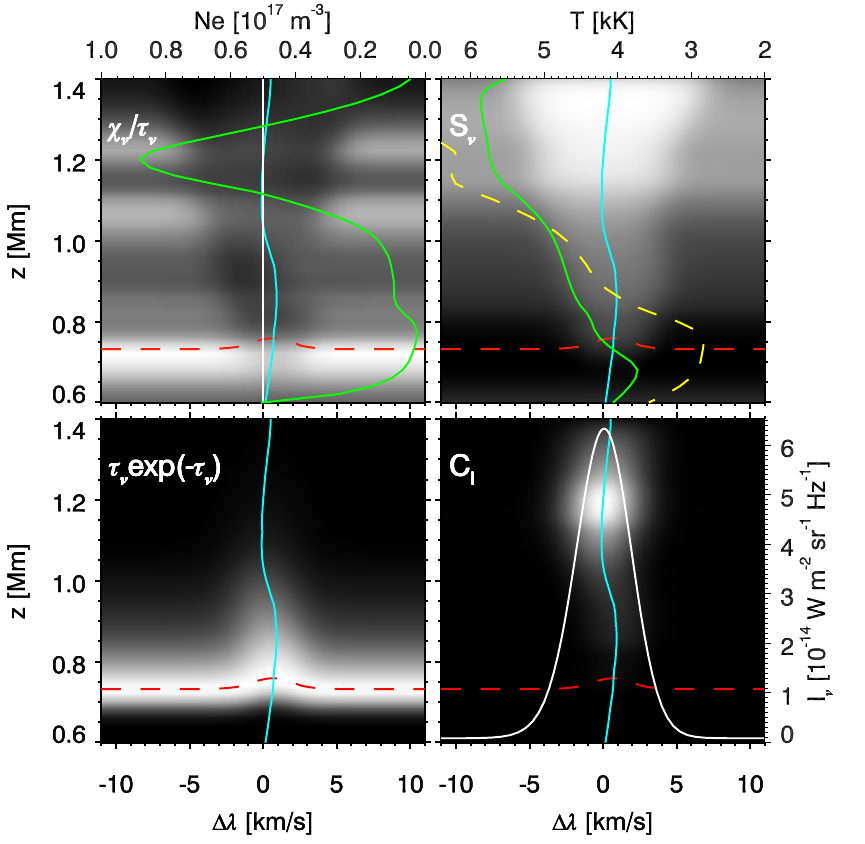} 
     \caption{Intensity formation for the \oil\ in a column of the 3D atmosphere characterised by 
     moderate velocities. 
     The label in the top-left corner of each panel gives the quantity shown as a grey-scale image 
     as function of wavelength (given as Doppler shift from the rest frequency, positive means redshift) and height $z$.  
     The $\tau$ = 1 height is given as a red dashed line and the vertical velocity as a blue solid line 
     in all four panels with a white straight line denoting the rest frequency in the upper left panel.
     The green solid line in the upper left panel gives the electron density with a scale at the top.
     In the upper right panel we show the temperature as a yellow dashed line and the source
     function at a frequency given by the velocity profile as a green line, both with the scale 
     given at the top.
     In the lower right panel we show the full contribution function (the product of the terms shown 
     as grey-scale images in the other three panels) together with the emergent intensity profile
     with a scale to the right.
     }
   \label{fig:oi_form1}
\end{figure}

\begin{figure} 
   \centering
     \includegraphics[width=\columnwidth]{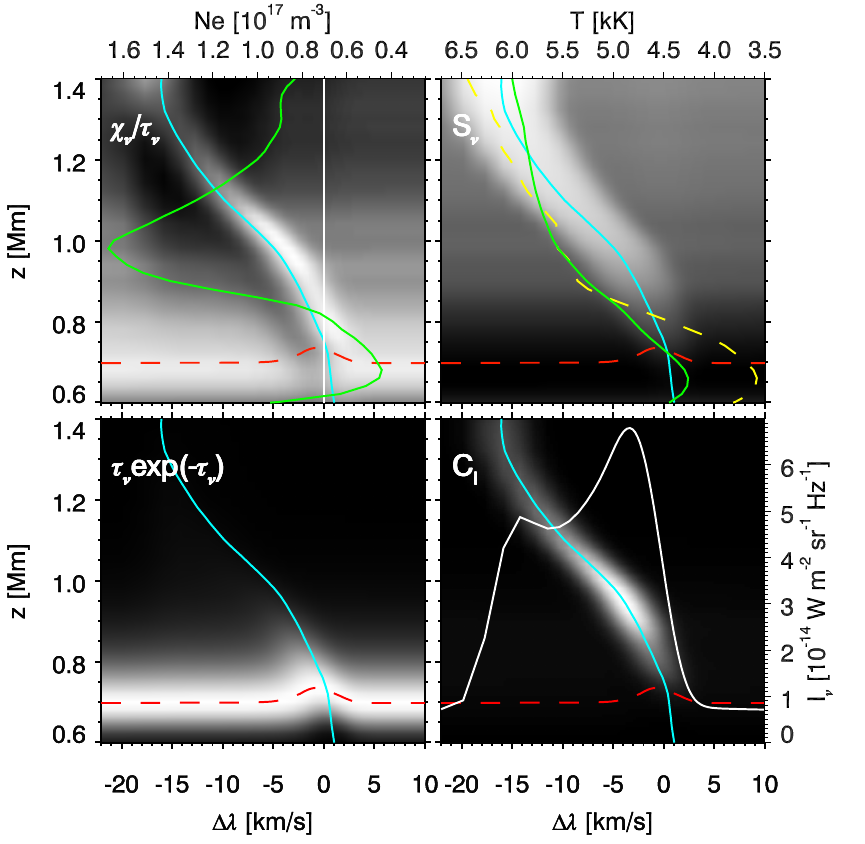} 
     \caption{As Figure~\ref{fig:oi_form1} but for a column with a large velocity amplitude through the 
     atmosphere($\sim$ 15\,\kms).  The line profile is highly asymmetric and the line is broadened
     by the velocity range within the \oi\ line forming region (0.7--1.4\,Mm in this case). 
     Note the expanded wavelength scale compared with Figure~\ref{fig:oi_form1}.
     }
   \label{fig:oi_form2}
\end{figure}

Figure~\ref{fig:oi_form1} shows the formation at an atmospheric column with moderate 
velocities. The continuum is formed at $\tau = 1$ located at a height of 0.73~Mm where the
temperature is 3\,kK but with a decoupled source function at a radiation temperature of 
4\,kK. The line core has
optical depth unity only slightly higher, at $z = 0.75$\,Mm, but the intensity is formed much higher up,
at $z = 1.2$\,Mm (shown by the contribution function distribution in the lower right panel
of Figure~\ref{fig:oi_form1}). The intensity comes from there because of the enhanced
source function at this height (upper right panel). The high source function comes from a peak
in the electron density (upper left panel). 
The line formation is thus optically thin in this case.
The profile is not much modified by the velocity field and
has a Gaussian shape with an 1/e width of 2.67\,\kms\ which should be compared with the
thermal width of 2.55\,\kms\ for the atmospheric temperature of 6.3\,kK at the formation height.

Figure~\ref{fig:oi_form2} shows the formation at a location where there 
is a substantial velocity field in the line forming region. The continuum
is formed at a height of 0.7\,Mm where the radiation temperature of the source function
is 4.5~kK. The continuum intensity is therefore higher than in
the case illustrated in Figure~\ref{fig:oi_form1}. The contribution function to intensity is significant
over a large part of the atmosphere with a low optical depth. The line is thus optically thin. The line profile gets broadened by the velocity gradients
in the line forming region, in the 0.7--1.4\,Mm height range. The 1/e width of the profile is
9.8\,\kms, substantially larger than the thermal width at the temperature of the line forming region.
This line-width - velocity profile relation can be used as a useful diagnostic of non-thermal broadening, see Section~\ref{sec:nthb}.
Note that the double-peaked line profile is caused by the velocity gradient and not by optically thick line formation.
 
\section{Diagnostic potential}\label{sec:diag}

\begin{figure*} 
   \centering
       	\includegraphics{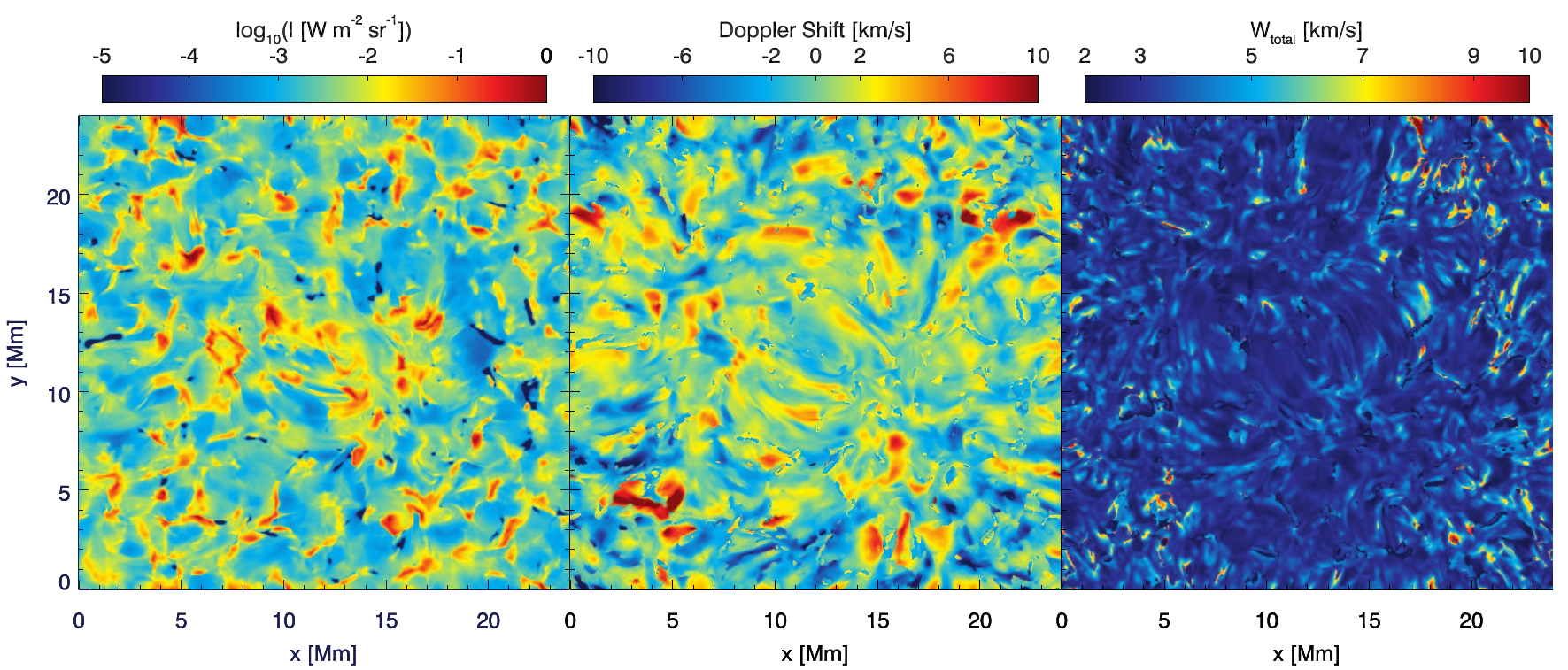} 
     \caption{\edt{Total intensity} (left), Doppler shift of the maximum emission (middle) and
     total line-width (right) of the \oil.
     }
   \label{fig:oi_3panel}
\end{figure*}

In this section we present synthetic profiles calculated from the 3D atmospheric snapshot and how observables are
related to physical quantities in the atmosphere, thereby exploring the diagnostic potential of the \oil. 
Figure~\ref{fig:oi_3panel} summarizes the properties of the synthetic \oil\ calculated throughout the 3D simulation box by
showing the total intensity, the Doppler shift of the maximum emission and the total line-width (given as 1/e width, see
Equation~\ref{eq:wtotal} for a definition). 
The total intensity is typically 2\,mW\,m$^{-2}$\,sr$^{-1}$ with a typical range of $0.3-65$\,mW\,m$^{-2}$\,sr$^{-1}$.
The Doppler shift is up to 10\,\kms. 
The total line-width is over a large part of the simulation box $2-4$\,\kms\ with maximum width up to 12\,\kms, typically in the 
corners of the simulation box where the magnetic field is the weakest. These locations have acoustic shocks propagating 
through the line-forming region, causing large line-widths.

Before we look at relations between observables and atmospheric properties, 
we inspect the contribution function to intensity to establish from where in the atmosphere observables are encoded.

\subsection{Contribution function to intensity}\label{sec:contri}

\begin{figure}
   \centering
     	\includegraphics[width=\columnwidth]{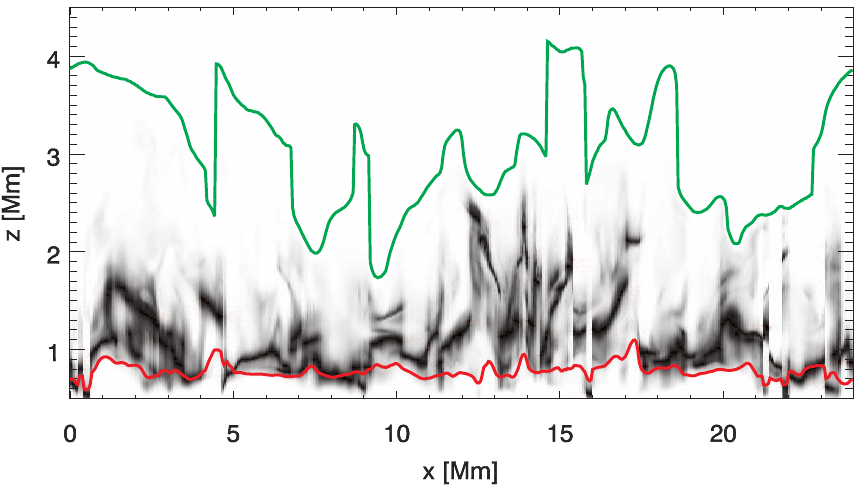} 
     \caption{Contribution function to the intensity of the \oi\ line core along a cut through the atmospheric model at $X=12$\,Mm
     together with the $\tau=1$ height of the line core (red solid line) and temperature contour at 30\,kK (green). The contribution
     function has been individually scaled for each column for increased visibility. Note that the Z-axis has been stretched.
     }
   \label{fig:o_contri}
\end{figure}

Typically the \oi\ line has a single emission peak without central reversal. 
We define the line core to be the location of maximum emission. Figure~\ref{fig:o_contri} shows the contribution function 
to intensity (Equation~\ref{eq:cinu}) at this core wavelength for
a cut through the 3D atmosphere at $x=12$\,Mm \citep[same cut as used in ][]{bhavna1, bhavna2}.

It is obvious that the intensity in the core of the \oil\ mostly comes from heights above the $\tau = 1$ height. The \oi\ line thus
shows optically thin formation. 

Correlations with atmospheric conditions are often made with the conditions at optical depth unity. For the \oil, with its optically 
thin formation, this is obviously a bad choice. Instead we define the formation height from the first moment of the 
contribution function:

\begin{equation}
z_{\rm fm} = \int_{-\infty}^{\infty} z\, C_{I_\nu}\!(z) dz.
\end{equation}

We correlate with mean atmospheric parameters calculated in the same fashion as
a contribution function weighted average:

\begin{equation}
V_{\rm fm} = \int_{-\infty}^{\infty} V(z)C_{I_\nu}\!(z) dz
\end{equation}

where $V(z)$ represent the physical quantity along a column of the atmosphere, e.g.\ the electron density, the temperature, or the vertical velocity, $C_{I_\nu}\!(z)$ is the contribution function to intensity and $V_{\rm fm}$ is the contribution function weighted
average of the physical quantity $V$.

\subsection{Velocity}\label{sec:v}

\begin{figure} 
   \centering
   \includegraphics[width=\columnwidth]{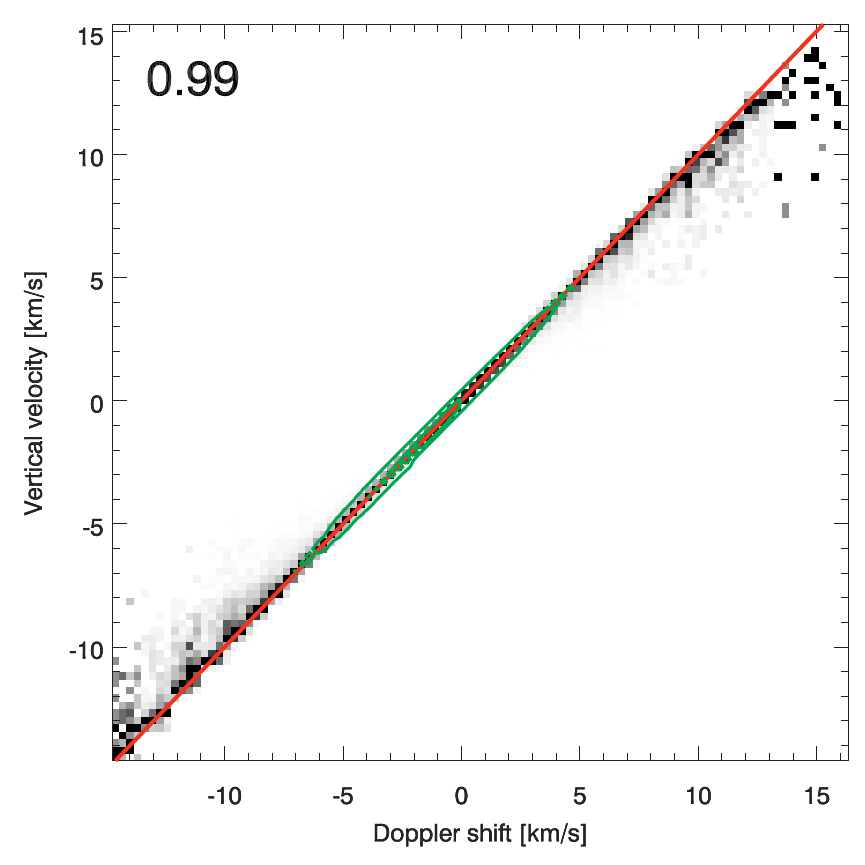}
   \caption{Probability Density Function (PDF) of the vertical velocity as a function of the Doppler shift of the line core.
   Each column is scaled to maximum contrast to increase visibility.
   The inner green contour encompasses 50\% of all points and the outer contour 90\%.  
   The Pearson linear correlation coefficient is given in the upper left corner. 
   The red line denotes the line y=x. 
    }
   \label{fig:vz_ds}
\end{figure}

Figure \ref{fig:vz_ds} shows that there is a tight correlation between the Doppler shift of the line core 
(defined to be the wavelength of maximum
emission) and the contribution function weighted vertical velocity. This makes the \oil\ a good velocity diagnostic. 
The shift of the \oi\ line is usually small (90\% of the data points lie within  $\pm$ 5 \kms) and the average shift is
1.5\,\kms. This average shift is caused by a global oscillation that is of this value at the time of the simulation snapshot.
The temporal average profile over a number of snapshots shows a shift very close to zero.
Using the \oil\ as a wavelength reference, as is done for \iris, is thus warranted.

\subsection{Electron density}\label{sec:diag_ne}

\begin{figure} 
   \centering
   \includegraphics[width=\columnwidth]{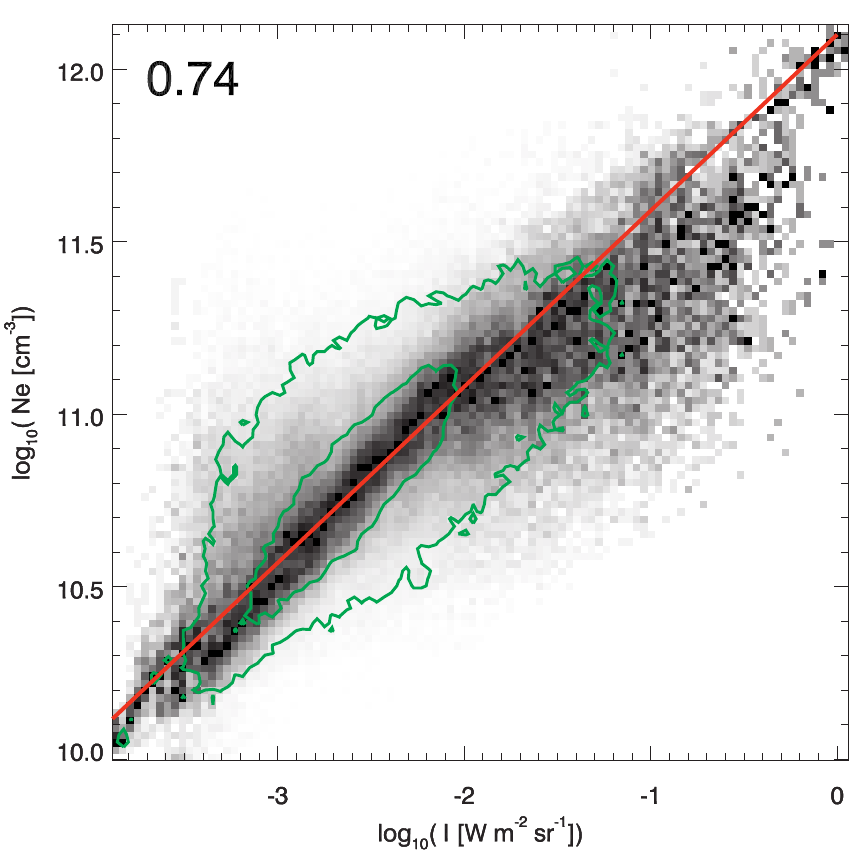}   
     \caption{PDF of the logarithm of the contribution function weighted electron density as a function of the logarithm of the \oi\ total line intensity. 
   Each column is scaled to maximum contrast to increase visibility.
   The green contours encompass 50\% and 90\% of all points.  
   The Pearson linear correlation coefficient is given in the upper left corner. 
   The red line denotes the line $\log_{10}{N_e} = 0.51 \log_{10}I +12.10$, which is equivalent to  $I \sim N_e^{1.97}$. 
   For more details see discussion in Sec. \ref{sec:diag_ne}.
   }
   \label{fig:ne_i}
\end{figure}

As discussed in Section~\ref{sec:steady}, the main channel of the \oi\ emission is through cascading from the continuum. 
The \oi\ ionization degree is proportional to the hydrogen ionization degree, hence the population of the \oi\ continuum will be 
proportional to the proton density. 
In the chromosphere, hydrogen is the dominant electron donor and the proton density can be well approximated by the electron density. 
The radiative recombination introduces another electron density dependency. 
We therefore anticipate the \oil\ total intensity to be proportional to the square of the electron density. 
Figure~\ref{fig:ne_i} shows the PDF of the logarithm of the  contribution function weighted electron density 
as a function of the logarithm of the \oi\ total line intensity. 
From the linear regression we get a slope 0.51 in the $\log_{10}{N_e}= f(\log_{10}I)$  plot, equivalent to
$I \sim N_e^{1.97}$. This confirms that the total intensity indeed is proportional to the electron density squared.
There is, however, a substantial width to the relation because of the temperature dependency of the recombination 
coefficient, see Figure~\ref{fig:ratecoef}.

\subsection{Non-thermal broadening of the \oi\ line profile}\label{sec:nthb}

In Section~\ref{sec:steady_mov} we saw that the \oi\ line-width can be sensitive to the velocity amplitudes in the
line formation region. Here we examine this relation further.

The total broadening, $W_{\rm total}$, thermal broadening $W_{\rm th}$, and non-thermal broadening $W_{\rm nth}$,  
are defined as follows:

\begin{eqnarray}
W_{\rm total} &=& \sqrt{W_{\rm FWHM}^2 \over 4 \ln (2)}, \label{eq:wtotal}\\
W_{\rm th} &=& \sqrt{2kT/m}, \label{eq:wth}\\
W_{\rm nth} &=& \sqrt{W_{\rm total}^2 -  {W_{th}}^2}, \label{eq:wnth}
\end{eqnarray}

where $k$, $m$, $T$,  $W_{\rm FWHM}$ denote the Boltzmann constant,  the atomic mass, the temperature and the full-width at half maximum from our \oi\ line profile, respectively.

The measure of the velocity profile $V(z)$ that is relevant to the line formation is defined as follows:


\begin{equation}\label{eq:vw}
V_\textrm{w} = \sqrt{2}  \ \frac{\int (\delta V(z))^2 C(z) dz}{\int C(z) dz}
\end{equation}

where 
\begin{eqnarray}
&&C(z) = \int_{\nu_b}^{\nu_r} C_{I_\nu}(z) d\nu,\\
 &&\delta V(z) = V(z)- V_{\rm core},  
\end{eqnarray}
 
where $V_{\rm w}$ is $\sqrt{2}$ times the weighted velocity standard deviation,  $\nu_b$ and $\nu_r$ are the blue- and the red-side 
frequency that defines the FWHM and $C_{I_\nu}(z)$, $V(z)$ and $V_{\rm core}$ are the contribution function to intensity, the vertical
velocity in the atmosphere and the Doppler shift of the defined line core. For a Gaussian velocity distribution, $V_{\rm w}$ would give
a 1/e broadening of the absorption profile of the same amount, which makes this definition appropriate for a direct comparison with
$W_{\rm nth}$ as defined above.

\begin{figure} 
   \centering
    	 \includegraphics[width=\columnwidth]{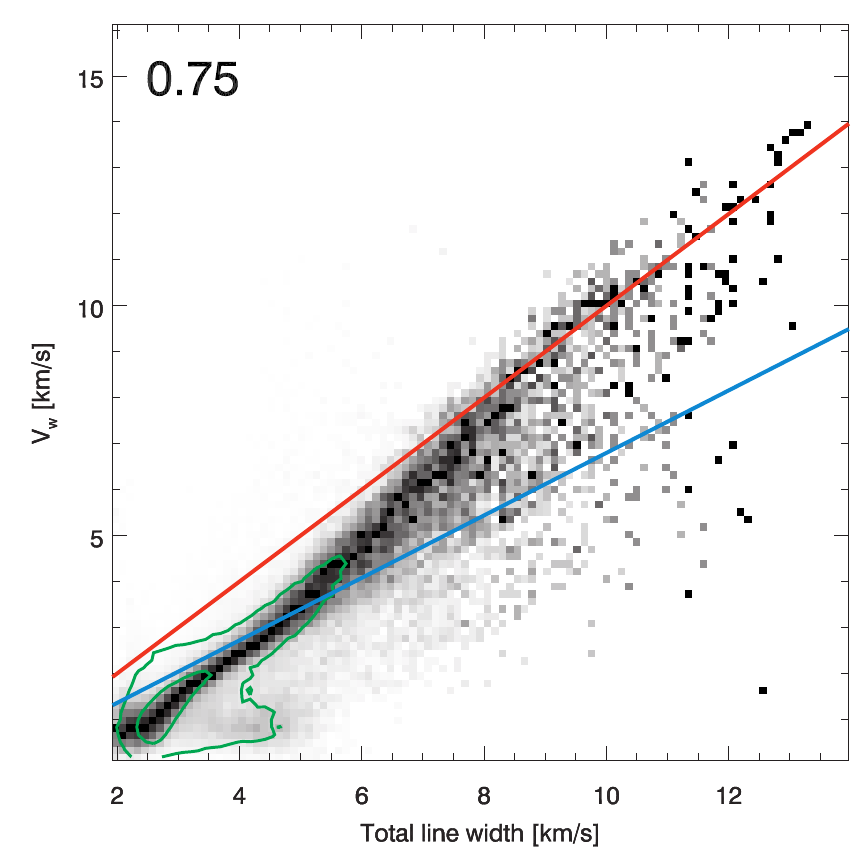} 
     \caption{PDF of $V_{\rm w}$ (Equation~\ref{eq:vw}) as a function of the total broadening, 
     $W_{\rm total}$ (Equation~\ref{eq:wtotal}).
   The green contours encompass 50\% and 90\% of all points.  
   The Pearson linear correlation coefficient is given in the upper left corner. 
   The diagonal red straight line denotes y = x, which is the \edt{expected relation} for a Gaussian velocity distribution; 
   the blue straight line denotes y = 0.68 x, which is the \edt{expected relation} for a flat velocity distribution.
}
   \label{fig:vw_wtot}
\end{figure}

\begin{figure} 
   \centering
    	 \includegraphics[width=\columnwidth]{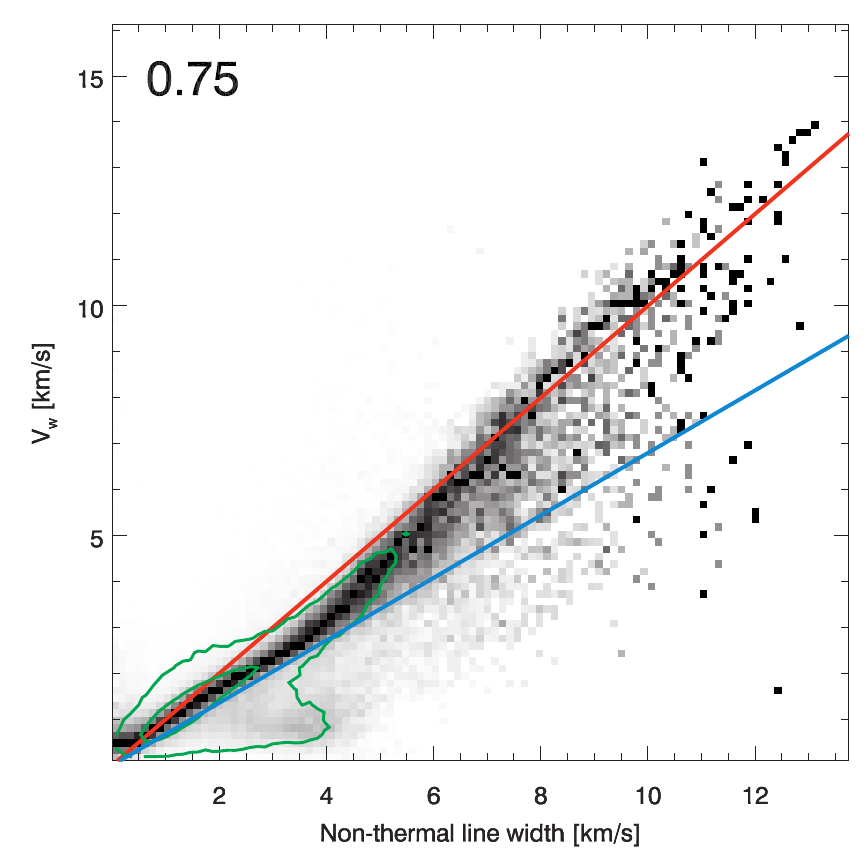} 
     \caption{PDF of $V_{\rm w}$ (Equation~\ref{eq:vw}) as a function of the non-thermal broadening, 
     $W_{\rm nth}$, of the line profile (Equation~\ref{eq:wnth}). 
   Each column is scaled to maximum contrast to increase visibility.
   The green contours encompass 50\% and 90\% of all points.  
   The Pearson linear correlation coefficient is given in the upper left corner. 
   The diagonal red straight line denotes y = x, which is the \edt{expected relation} for a Gaussian velocity distribution; 
   the blue straight line denotes y = 0.68 x, which is the \edt{expected relation} for a flat velocity distribution.
   }
   \label{fig:vw_wnth}
\end{figure}

Figure~\ref{fig:vw_wtot} shows the PDF of the measure $V_{\rm w}$ as a function of the total line-width of the intensity profile of the
\oil. The relation expected for a Gaussian velocity distribution without thermal broadening is shown with a straight red line. The
relation expected for a flat velocity distribution (all the velocities are equally probable, e.g., a linear velocity gradient) 
is shown with a blue straight line.
It is clear that when the total line-width is small ($\leq$ 6 \kms), the total line-width is larger than the velocity 
amplitudes indicate because the broadening is dominated by thermal broadening rather than non-thermal broadening. 
As the line-width grows bigger the broadening gets more and more dominated by non-thermal broadening.

With the knowledge of the temperature structure in the atmosphere, we may directly correlate the velocity field with
the non-thermal broadening (using Equation~\ref{eq:wnth}). Figure~\ref{fig:vw_wnth} shows this correlation together with
the expectations from a Gaussian and a flat velocity distribution.
The correlation from our simulation shows a tendency of a flat distribution 
when the non-thermal broadening is small ($\leq $4 \kms),  and it gradually approaches a Gaussian distribution for larger non-thermal
broadening. Thanks to the optically thin formation, the \oi~133.56\,nm line-width is a very good diagnostic of non-thermal broadening.

\section{Comparison with observations}\label{sec:obs}

\begin{figure} 
   \centering
    \includegraphics[width=\columnwidth]{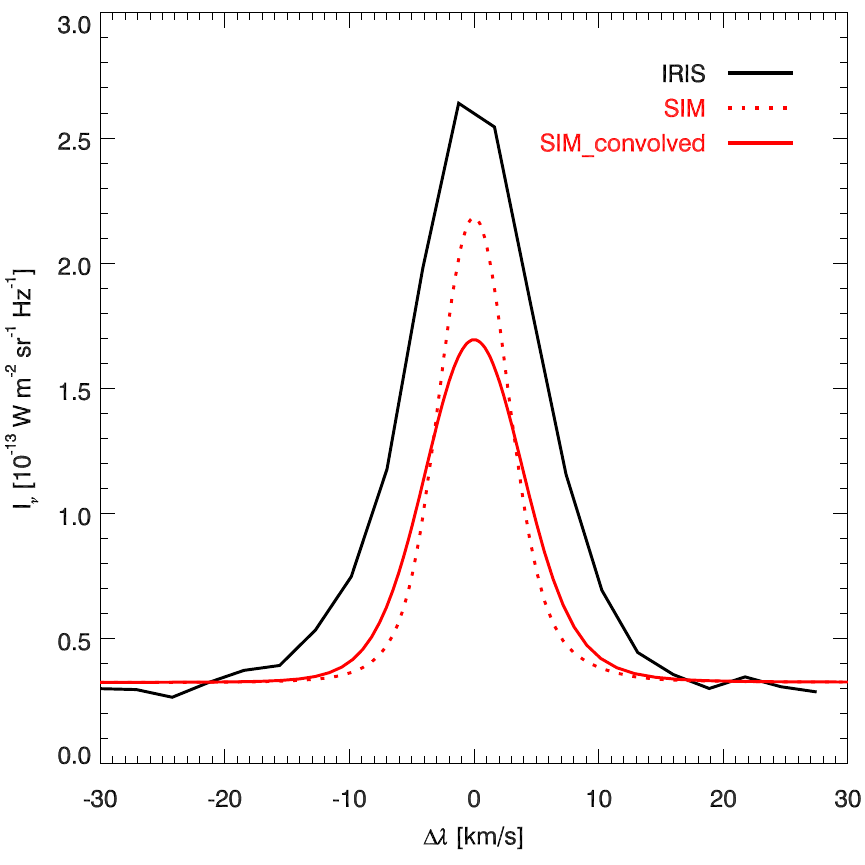} 
     \caption{Comparison of the average
synthetic spectrum from our simulation convolved with the \iris\ PSF (red) with observations of the
quiet sun at disk center by  \iris\ from 2014 \edt{August 10 at 03:12:39} UT (black).
 }
   \label{fig:sim_obs}
\end{figure}

We have explored correlations between atmospheric parameters in a snapshot of a radiation-MHD simulation of the solar atmosphere
and observables of the \oil. A relevant question is how well the simulated line-profile reproduces the real, observed, intensity profile.
Figure~\ref{fig:sim_obs} shows the average synthetic \oil\ profile compared with the average profile from an \iris\ observation. 
The synthetic profile has been convolved with the \iris\ spectral point-spread-function 
\citep[FWHM of 26\,m\AA,][]{2014SoPh..289.2733D}.
The \iris\ data were acquired on 2014 August 10 starting at 03:12:39 UT and comprise a 400-step raster of a rather quiet area close 
to disk center. The exposure time was 30\,s. 
The standard \iris\ wavelength calibration is based on the \oil\ such that the mean Doppler-shift is zero by definition. 
The "continuum" signal is dominated by contamination light from longer wavelengths. 
We have therefore subtracted this constant level from the observed profile and added the continuum level from the simulations.
The synthetic profile is more narrow than the observed profile. 
This deficiency of the simulation is also seen in profiles of the \ion{Mg}{2}~k line
\citep{mg1,mg2,mg3} and the \ion{C}{2} multiplet at 133.5\,nm \citep{bhavna1,bhavna2}. Given the good correlation between 
the \oil\ non-thermal width and the velocity field in the atmosphere this is an indication that the simulation has too small 
macroscopic velocities.

\section{Conclusion}\label{sec:concl}

In this work we have studied the basic formation mechanism of the \oi\ 135.56\,nm intersystem line. 
The line shows an optically thin formation and the emission gets contributions from the whole chromosphere. 
The radiative rate is dominated by a recombination cascade
and the total intensity is therefore proportional to the electron density and the population density of ionized oxygen. The ionization
balance is dominated by charge transfer collisions with hydrogen such that the final total intensity is proportional to the 
electron density squared.

The Doppler shift of the \oil\ is very closely related to the velocity averaged over the formation region.

Due to the optically thin nature of the \oil\ formation we get a line-width that is a direct measure of the thermal and non-thermal
broadening. 
The thermal width is rather small (3.2\,\kms\ at a temperature of 10~kK, which is a temperature at the upper range of what
we find) such that a profile with an 1/e width of more than
5 \kms\ is dominated by non-thermal broadening. 
The \oil\ thus provides a very good measure of non-thermal velocities in 
the chromosphere and is an excellent complement to the optically thick \iris\ chromospheric diagnostics 
such as the \ion{Mg}{2} h \& k lines and the \cii\ lines around 133.5\,nm.

\begin{acknowledgements}
The research leading to these results has received funding from the European Research Council under the
European Union's Seventh Framework Programme (FP7/2007-2013) / ERC grant agreement no 291058.
This research was supported by the Research Council of Norway through the grant "Solar Atmospheric 
Modelling" and through grants of computing time from the Programme for Supercomputing. 
\end{acknowledgements}

\end{document}